\documentclass[
    aps,
    prapplied,
    reprint,
    superscriptaddress,
    amsmath,
    amssymb
    ]{revtex4-2} 
\usepackage{graphicx}
\usepackage{bm}
\usepackage{hyperref}
\usepackage{xcolor}
\usepackage{siunitx}
\DeclareSIUnit{\ions}{ions}
\newcommand{\Rn}{R_{\mathrm{n}}}
\newcommand{\Ic}{I_{\mathrm{c}}}
\newcommand{\Tc}{T_{\mathrm{c}}}
\newcommand{\IcRn}{I_{\mathrm{c}}R_{\mathrm{n}}}
\newcommand{\ET}{E_{\mathrm{T}}}
\newcommand{\YBCO}{YBa$_{2}$Cu$_{3}$O$_{7-\mathrm{\delta}}$}
\newcommand{\YBCOseven}{YBa$_{2}$Cu$_{3}$O$_{7}$}
\newcommand{\lambdazeroIc}{\lambda_{0}^{(\Ic)}}
\newcommand{\lambdazeroRn}{\lambda_{0}^{(\Rn)}}
\newcommand{\lambdazeroIcRn}{\lambda_{0}^{(\IcRn)}}
\newcommand{\IV}{\textit{I-V}}
\begin{document} 
\title{Josephson Transport in \texorpdfstring{\YBCOseven}{YBa2Cu3O7} weak links created by focused-helium-ion-beam irradiation: Analysis based on diffusive-SNS-junction model} 
\author{Tetsuro~Misawa} 
\thanks{tetsuro-misawa@aist.go.jp}
\author{Shigeyuki~Ishida} 
\author{Hiroshi~Eisaki} 
\author{Yukinori~Morita} 
\author{Shinichi~Ogawa} 
\author{Chiharu~Urano} 
\thanks{c-urano@aist.go.jp}

\affiliation{National Institute of Advanced Industrial Science and Technology, Tsukuba 305-8563, Japan} 
\begin{abstract}
Fabrication of YBCO weak links by focused helium ion beam irradiation is a promising approach for realizing high-temperature superconducting Josephson junction devices.
Although empirical dose-characteristic relationships have been established, the underlying transport mechanisms remain unclear.
In this study, we perform a detailed investigation of the transport properties of YBCO weak links fabricated using a helium ion microscope (HIM) and provide a unified phenomenological description of the observed behavior based on the theory of SNS junctions with a diffusive metallic interlayer.
We demonstrate that the temperature dependence of the critical current $\Ic$ and the $\IcRn$ product are well described by diffusive SNS junction models over a wide temperature range.
Analyses show that the observed dose dependences of $\Ic$ and $\IcRn$ cannot be explained solely by variations in the effective Thouless energy $\ET$.
The discrepancy suggests reduced interface transparency and a reduction in the density of states, leading to a decrease in the effective number of conducting channels contributing to transport.
This interpretation is also consistent with the observed exponential increase in $\Rn$ with irradiation dose.
These results provide a diffusion-based framework for understanding Josephson transport and guiding junction design in helium-ion-irradiated YBCO weak links.
\end{abstract} 
\maketitle 
\section{Introduction}
Superconductivity provides a platform for circuit devices characterized by ultralow dissipation, ultrafast operation, and low power consumption, enabling functionalities and performance beyond those attainable with conventional semiconductor technologies~\cite{Bairamkulov2024}.
In addition, the macroscopic quantum coherence inherent in the superconducting state has led to widespread applications in precision metrology, including highly sensitive magnetic sensors and quantum voltage standard technologies~\cite{Hamilton2000,Benz1996,Zhou2026,Khorshev2019}.
Because the Josephson junction constitutes the fundamental element of virtually all superconducting electronic systems, the realization of Josephson junctions with properties optimized for specific applications and operating conditions remains one of the most important challenges in contemporary superconducting electronics.

Josephson junction fabrication technology has reached a high level of maturity for elemental superconductors such as Al and Nb~\cite{Tolpygo2015,Tolpygo2016}, as well as simple compound superconductors including NbN~\cite{Radparvar1987}, enabling highly reproducible device fabrication and large-scale integration.
In contrast, high-temperature superconducting (HTS) Josephson junctions continue to face technological challenges, such as the precise control of junction characteristics and the reduction of parameter variability, which have hindered their integration into large-scale circuits.
The coherence length of cuprate HTS materials is typically on the order of only a few nanometers.
Together with the anisotropic nature of their superconducting order parameter, this leads to a high sensitivity of junction characteristics to microscopic crystal defects and to the nanometer-scale roughness of the junction interfaces~\cite{Hilgenkamp2002}.
Moreover, in HTS materials, local superconducting order parameter and low-energy electronic states are strongly influenced by carrier concentration, oxygen ordering, and crystalline disorder.
Consequently, the inherent complexity of the underlying material physics poses major challenges for the realization of reproducible device fabrication processes.

Helium ion microscopy (HIM)-based fabrication has emerged as a promising approach for realizing Josephson junctions in cuprate high-temperature superconductors~\cite{Cybart2015}.
In this method, a superconducting weak link is formed by locally suppressing superconductivity in thin films of materials such as \YBCO (YBCO) using a helium ion beam focused to a sub-nanometer scale.
The junction properties can be continuously tuned over a wide range, from SIS-like to SNS-like characteristics, by adjusting the helium ion irradiation dose.
In addition, the ability to directly write weak-link structures into superconducting circuits provides appreciable design flexibility and offers major advantages for the large-scale integration of superconducting devices.

Josephson junctions realized by this technique have been studied in a variety of superconducting materials including MgB${}_{2}$,~\cite{Kasaei2018} Bi2212,~\cite{Wang2021} and Nb-based compounds~\cite{Ruhtinas2025,Li2023}, with YBCO attracting particular attention.~\cite{Cybart2015,Cho2018,Couedo:2020aa,Chen2022}
The high sensitivity of YBCO to helium ion irradiation enables substantial modification of its physical properties and, consequently, tuning of the junction characteristics over a wide parameter range.
Extensive investigations have therefore been conducted on the dependence of the critical current $\Ic$, normal state resistance $\Rn$, and characteristic product $\IcRn$ on the irradiation dose~\cite{Muller2019,Chen2022,Karrer2024,Ogawa2024,Parachikunnumal2026}.
Among these studies, M\"uller et al. reported an empirical relationship whereby $\Ic$ and $\Ic\Rn$ decrease exponentially with increasing helium ion dose, while $\Rn$ exhibits an exponential increase.

Despite the recent progress achieved in HIM-fabricated Josephson junctions, especially in their applications to SQUIDs, digital circuits and other devices, a fundamental understanding of the physical properties and transport phenomena of helium-ion-induced weak links has yet to be established.
In cuprate high-temperature superconductors such as YBCO, superconducting order and low-energy electronic states are highly sensitive to oxygen ordering, and crystalline disorder.~\cite{Liang2006,Alloul2009}
Consequently, helium-ion irradiation may do more than simply introduce structural defects; it can result in changes in carrier density, carrier diffusion constant, density of states within the weak-link region simultaneously~\cite{Muller2019,Zaluzhnyy2024}.
Establishing the relationship between the transport characteristics of these weak links and the underlying nanoscale material properties is therefore of fundamental importance.
Such knowledge is indispensable for developing Josephson junction technologies based on YBCO and other high-temperature superconductors on a physical foundation rather than on empirical knowledge alone, and for advancing HIM-based junction fabrication toward a truly physics-driven engineering methodology.

In this work, we investigate the transport properties of YBCO weak links fabricated by a HIM through a phenomenological analysis based on a physically motivated transport model.
Specifically, the transport characteristics are analyzed within the framework of dirty-limit SNS junction theory, where the weak-link region is modeled as a diffusive metallic interlayer.
This approach enables the observed transport behavior to be described in terms of a limited set of phenomenological parameters, such as the effective junction length, diffusion coefficient, and interface transparency.
In addition, the model provides a means of extracting information on the nanoscale physical properties of YBCO in the irradiated weak-link region from transport measurements.
By establishing relationships between the transport properties and the underlying physical characteristics of YBCO, we aim to clarify how nanoscale material characteristics manifest themselves in Josephson transport.
More broadly, this work explores the potential of transport measurements not only as a tool for device characterization but also as a probe of microscopic electronic properties in cuprate high-temperature superconductors.

\section{Methods}
In this study, weak-link devices were fabricated from YBCO thin films purchased from Ceraco ceramic coating GmbH.
The films were deposited directly on MgO (100) substrates without the use of a buffer layer.
A 25-nm-thick YBCO (001) film was grown, with the YBCO (100) and (010) directions aligned parallel to the MgO (100) substrate orientation.
X-ray diffraction (XRD) measurements confirmed that the YBCO films were \textit{c}-axis oriented.
Furthermore, phi-scan measurements revealed a fourfold-symmetric pattern of the (038) reflection, which indicates that the YBCO films used in this study are twinned films containing both \textit{a}-axis and \textit{b}-axis-oriented domains within the \textit{ab}-plane.
Following the deposition of the YBCO layer, a 100-nm-thick Au film was deposited \textit{in situ}.
This Au layer served both as a protective cap for the YBCO film and as a bonding pad for device wiring.
The deposited YBCO films exhibited a superconducting transition temperature $\Tc$ of \qty{84.3}{\kelvin} and a critical current density of \qty{2.8}{\mega\ampere/\cm^{2}} at \qty{77}{\kelvin}.

Using photolithography and low-pressure Ar ion milling, the YBCO films were patterned into bar-shaped four-terminal devices with a width of \qty{4}{\micro\meter}.
The devices were designed such that the current flowed parallel to either the crystallographic \textit{a}-axis or \textit{b}-axis of YBCO.
Subsequently, only the Au layer at the intended helium-ion irradiation site was removed by Ar ion milling to expose the underlying YBCO surface.

To form the weak links, a focused helium ion beam was irradiated across the center of each fabricated YBCO bar.
Helium ion irradiation was performed using a Zeiss ORION helium ion microscope.
The ion acceleration voltage was set to \qty{30}{\kilo\volt}, and the irradiation pitch was \qty{0.153}{\nm}.
The helium ion beam was focused to the smallest possible diameter by adjusting the focus using reference marks patterned on the substrate surface.
Based on the manufacturer's specifications, the beam waist was estimated to be approximately \qty{0.5}{\nm} under these conditions.
To avoid unnecessary exposure of the YBCO bars to helium ions, the irradiation positions were determined exclusively by referencing pre-fabricated alignment marks on the substrate, without direct imaging of the YBCO bar regions.
This procedure minimized unintended helium ion irradiation and thereby reduced irradiation-induced damage to the samples.
During irradiation, the entire sample was electrically grounded to prevent sample charging.
Irradiation was carried out on multiple YBCO bars fabricated on the same substrate to minimize the effect of nonuniformity in YBCO local composition.
The irradiation dose was varied from \qty{153}{\ions/\nm} to \qty{421}{\ions/\nm}.

\section{Results}
\subsection{Transport measurement and \IV characteristics}
\begin{figure*}[t]
    \centering
    \includegraphics[width=0.9\textwidth]{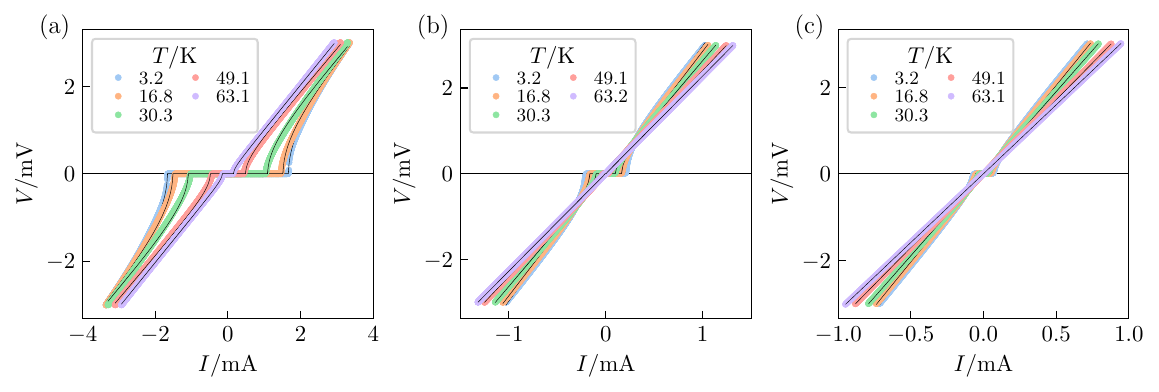}
    \caption{
        Current-Voltage characteristics of YBCO Josephson junctions at selected temperatures.
        The helium ion irradiation doses are (a) \qty{268}{\ions/\nm}, (b) \qty{382}{\ions/\nm}, (c) \qty{421}{\ions/\nm}, respectively.
        The black solid lines are fits to the RSJ model.
    }
    \label{fig:IVC}
\end{figure*}
The fabricated Josephson junctions were cooled and characterized using our cryogenic measurement system.
The samples were mounted in a pulse-tube cryocooler, and their current-voltage (\IV) characteristics and microwave-response properties were measured over the temperature range from \qty{3.2}{\kelvin} to \qty{75}{\kelvin}.
Electrical transport measurements were carried out using a Keithley 6221 current source and a Keithley 2182A nanovoltmeter.
A measured sample was selected using a switching box installed at room temperature, allowing each bar on a substrate to be addressed sequentially without warming the system.

Figures~\ref{fig:IVC}(a)-\ref{fig:IVC}(c) present the \IV characteristics measured at selected temperatures for weak links fabricated with helium-ion doses of \qtylist{268;382;421}{\ions/\nm}, respectively.
No hysteresis is observed in any of the devices, demonstrating overdamped transport characteristics.
In some of the fabricated weak links, the measured \IV curves exhibited flux-flow-like characteristics featured by a pronounced rounding near the onset of voltage rise, suggesting the persistence of superconducting pathways through the nominal weak-link region.

Therefore, in the present study, weak links exhibiting rounded \IV characteristics were excluded from the subsequent analysis.
Accordingly, the following analysis is restricted to the three weak links irradiated with doses of \qtylist{268; 382; 421}{\ions/\nm}, all of which exhibit a distinct and sharp voltage switching behavior.

The \IV characteristics of these weak links were fitted using the resistively shunted junction (RSJ) model.
For all investigated temperatures, the fitting residuals were small, indicating that the measured transport characteristics are well described by the RSJ model.
The weak links fabricated in this study exhibited only a small excess current even at low temperatures, making them well suited for investigating the temperature dependence of the junction properties.
For comparison, fitting was also performed using the resistively and capacitively shunted junction (RCSJ) model, which accounts for the junction capacitance.
For the device irradiated with \qty{268}{\ions/\nm}, the Stewart-McCumber parameter was found to reach a maximum value of approximately $0.2$ at low temperatures.
In addition, the $\IcRn$ product obtained from the RCSJ fitting was approximately \qty{3}{\percent} larger than that extracted using the RSJ model.
However, the RCSJ fitting became unstable at higher temperatures; in the vicinity of $\Tc$, the fitting procedure yielded values of $\Ic$ that increased with increasing temperature, which is clearly unphysical.
This behavior indicates that the additional fitting parameters introduced in the RCSJ model lead to overfitting of the experimental data.
For these reasons, the RSJ model was adopted consistently throughout the present study.

The black solid lines in Figs.~\ref{fig:IVC}(a)-\ref{fig:IVC}(c) correspond to fits to the RSJ model.
From these fits, $\Ic$ and $\Rn$ were determined for each device and temperature.

\begin{figure}[t]
    \centering
    \includegraphics[width=.9\columnwidth]{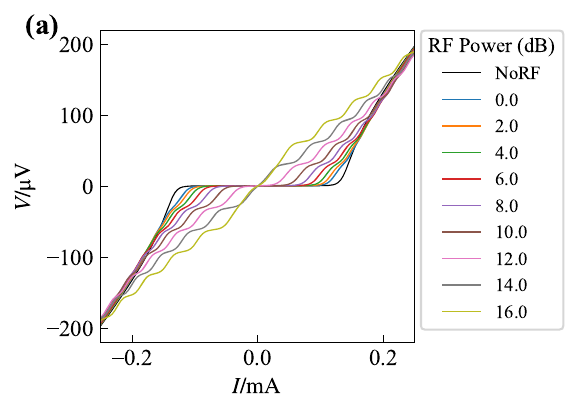}

    \includegraphics[width=.9\columnwidth]{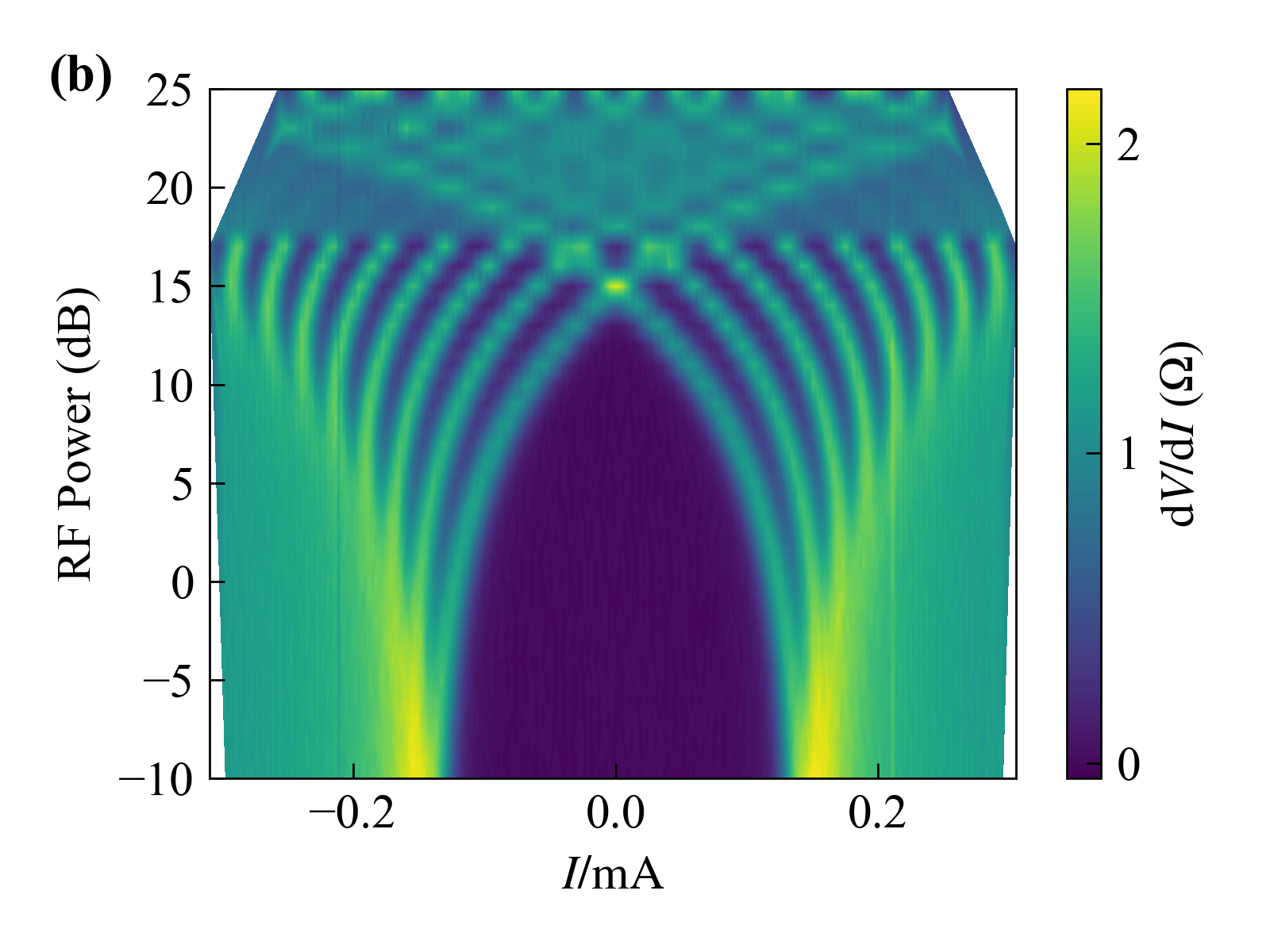}
    \caption{
        Shapiro steps observed at \qty{63.1}{\kelvin} in a YBCO Josephson junction fabricated with a helium-ion dose of \qty{268}{\ions/\nm}. The microwave frequency is \qty{15}{\GHz}.
        (a) Current-voltage characteristics under microwave irradiation. (b) Colormap of differential resistance.
        RF power is expressed in dB relative to \qty{1}{\milli\watt}.
        In the high-power range ($>\qty{17}{\dB}$), the colormap is obscured, which may be attributable to unexpected reflection of microwave in the measurement RF line system.
    }
    \label{fig:Shapiro}
\end{figure}
The formation of Josephson junctions in the fabricated weak links was verified through their high-frequency response under microwave irradiation.
Microwave signals were generated using an Agilent E8257D signal generator.
A quarter-wavelength antenna was constructed from the center conductor of a microwave coaxial cable and positioned in close proximity to the sample to irradiate the weak links with microwaves.
Figure~\ref{fig:Shapiro}(a) shows the \IV characteristics measured for the weak link fabricated with an irradiation dose of \qty{268}{\ions/\nm} under microwave irradiation at \qty{63.1}{\kelvin}.
The microwave frequency was \qty{15}{\giga\hertz}, and the output power of the signal generator was varied up to \qty{25}{\dB} (rereference: \qty{1}{\milli\watt}).
For all microwave power levels, plateau-like structures were observed at voltages corresponding to integer multiples of $hf/2e$.
These features are identified as Shapiro steps arising from the ac Josephson effect.
Figure~\ref{fig:Shapiro}(b) presents a color map of the differential resistance as a function of bias current and microwave power.
At microwave powers above \qty{17}{\dB}, unwanted effects such as microwave reflections became significant, resulting in a deterioration of the Shapiro-step pattern.
Shapiro steps were observed in all weak links fabricated with irradiation doses of \qtylist{268;382;421}{\ions/\nm}, confirming that Josephson junctions had been successfully formed in each device.

\subsection{Temperature and irradiation dose dependences of parameters}
\begin{figure*}[t]
    \centering
    \includegraphics[width=.95\textwidth]{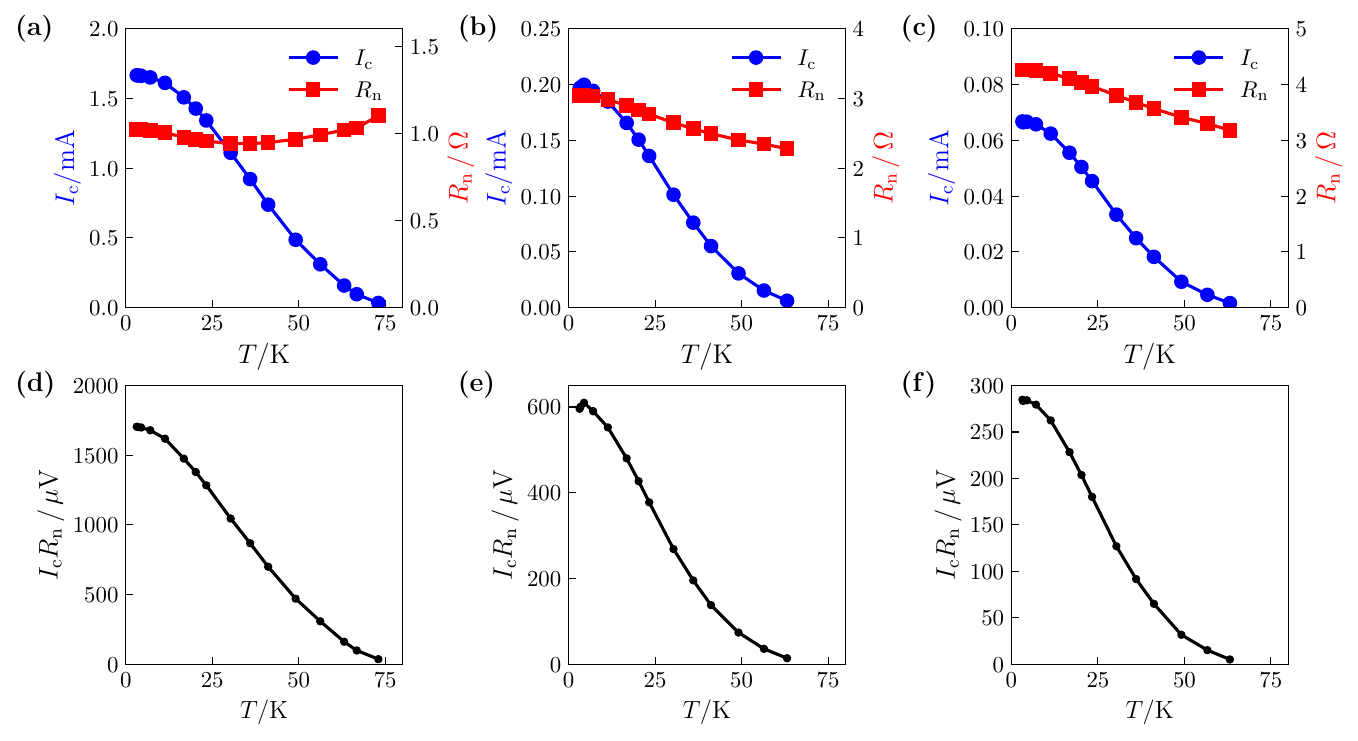}
    \caption{
        Temperature dependences of characteristic parameters for YBCO weak links.
        (a)-(c) Josephson critical current (blue circles) and normal resistance (red squares) for YBCO weak links fabricated with different helium ion doses. (d)-(f) $\IcRn$ product.}
    \label{fig:temp_dependence}
\end{figure*}
Figure~\ref{fig:temp_dependence} shows the temperature dependences of$\Ic$ and $\Rn$, which were extracted by fitting the current-voltage characteristics of the weak links.
For all devices, $\Ic$ increased monotonically with decreasing temperature.
The rate of increase gradually diminished below approximately \qty{10}{\kelvin}, and $\Ic$ approached a nearly constant value, indicating saturation at low temperatures.
The critical current densities at the lowest measurement temperature of \qty{3.2}{\kelvin} were \qtylist{1.7;0.20;0.067}{\mega\ampere/\cm^{2}} for the weak links irradiated with \qtylist{268;382;421}{\ions/\nm}, respectively.
All weak links exhibited only a weak temperature dependence of the normal-state resistance $\Rn$.
Although the devices irradiated with higher doses of \qtylist{382; 421}{\ions/\nm} showed a gradual increase in $\Rn$ at low temperatures, the overall variation was limited to approximately \qty{40}{\%} within the measurement temperature range.
Previous studies have reported that, at sufficiently high helium ion irradiation doses, the weak-link region becomes insulating and the resistance diverges at low temperatures~\cite{Muller2019}.
The weak links investigated in the present study exhibit resistance values that tend to approach finite constants as the temperature decreases.
This behavior suggests that the irradiated weak-link region remains metallic, albeit with a large residual resistance.

Among these samples, the temperature dependence of $\Rn$ is almost absent in the \qty{268}{\ions/\nm} sample.
Under the assumption that the irradiated YBCO in the weak-link region is close to the metal-insulator crossover regime, it is possible to estimate the effective junction length $L$.
According to previous studies in which crystalline disorder was introduced into YBCO by a large area helium-ion irradiation, the resistivity at the metal-insulator crossover is approximately \qty{800}{\micro\ohm\cm}~\cite{keppert_temporal_2024}.
Assuming that the resistivity of the \qty{268}{\ions/\nm} device is comparable to this value, $L$ can be estimated from the relation $\Rn=\rho L/A$ where $A$ is the cross-sectional area of the weak link.
This yields an effective length of approximately \qty{15}{\nm}.
This estimate suggests that the effective weak-link length $L$ can be substantially larger than the diameter of the focused helium ion beam itself.

Figures~\ref{fig:temp_dependence}(d)-(f) show the $\IcRn$ products calculated from the extracted values of $\Ic$ and $\Rn$.
At the lowest measurement temperature of \qty{3.2}{\kelvin}, the \mbox{$\IcRn$} products were \qtylist{1.7;0.60;0.28}{\milli\volt} for the devices irradiated with doses of \qtylist{268;382;421}{\ions/\nano\meter}, respectively.
As discussed above, the temperature dependence of $\Rn$ is relatively weak; therefore, the temperature dependence of the $\IcRn$ product is governed primarily by the variation of $\Ic$.

\begin{figure}[t]
    \centering
    \includegraphics[width=0.48\columnwidth]{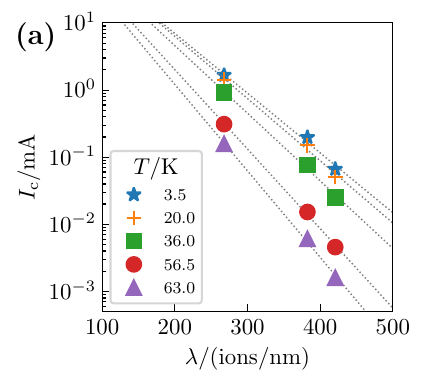}
    \includegraphics[width=0.48\columnwidth]{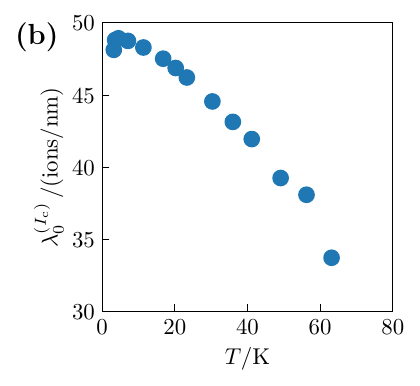}

    \includegraphics[width=0.48\columnwidth]{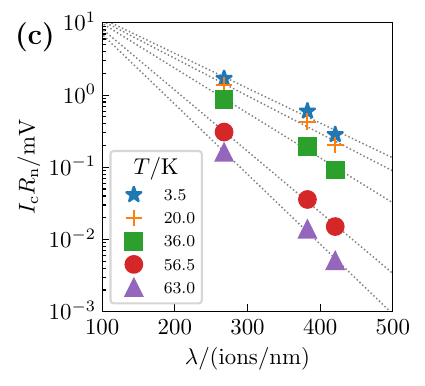}
    \includegraphics[width=0.48\columnwidth]{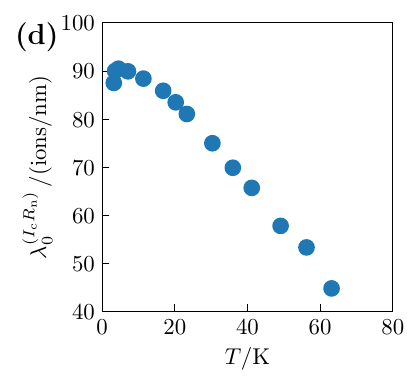}

    \includegraphics[width=0.48\columnwidth]{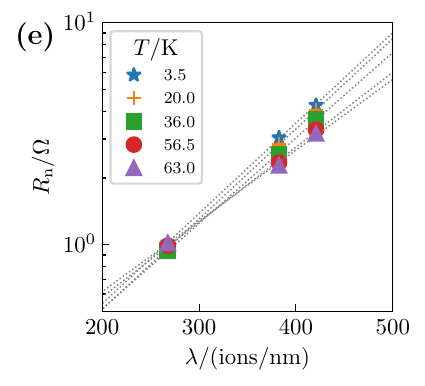}
    \includegraphics[width=0.48\columnwidth]{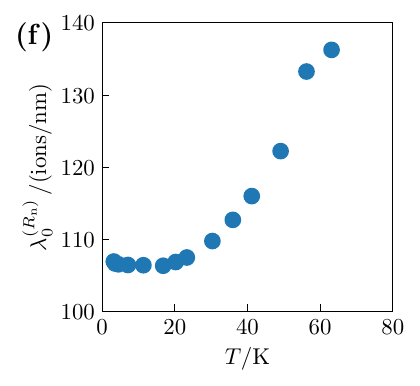}
    \caption{
        Dose ($\lambda$) dependence of (a) $\Ic$, (c) $\IcRn$, and (e) $\Rn$ for selected temperatures.
        The black dotted lines are fitting functions defined in the body.
        Temperature dependence of characteristic doses: (b) $\lambdazeroIc$, (d) $\lambdazeroIcRn$, and (f) $\lambdazeroRn$.
        }
    \label{fig:dose_dependence}
\end{figure}

The dose dependence of $\Ic$ at selected temperatures is shown in Fig.~\ref{fig:dose_dependence}(a).
At all temperatures, $\Ic$ decreases exponentially with increasing irradiation dose, in agreement with previous reports~\cite{Muller2019}.
The experimental data were fitted using the empirical relation $\Ic(\lambda)=\Ic^{(0)}\exp{}\left(-\lambda/\lambdazeroIc\right)$, where $\lambda$ is the irradiation line dose and $\lambdazeroIc$ is the characteristic dose. From this fitting procedure, the characteristic dose $\lambdazeroIc$ was obtained.
As shown in Fig.~\ref{fig:dose_dependence}(b), $\lambdazeroIc$ becomes larger at lower temperatures.
For example, $\lambda_{0}$ was \qty{49}{\ions/\nm} at \qty{3.6}{\kelvin}, whereas it decreased to \qty{34}{\ions/\nm} at \qty{63.1}{\kelvin}.
The value obtained at \qty{3.6}{\kelvin} lies between the characteristic doses previously reported for YBCO films on LSAT substrates (\qty{38}{\ions/\nm}) and those on STO and MgO substrates (\qty{130}{\ions/\nm})~\cite{Muller2019}.
The observation that $\lambdazeroIc$ exhibits a temperature dependence has important practical implications.
Specifically, it indicates that the precision required for irradiation-dose control in order to achieve uniform Josephson-junction characteristics depends on the intended operating temperature of the YBCO weak links.
At higher operating temperatures, $\lambdazeroIc$ is smaller, implying that fluctuations in irradiation dose during helium-ion exposure may have a stronger impact on the resulting junction properties.
Under these conditions, the larger value of $\lambdazeroIc$ reduces the sensitivity of the junction characteristics to variations in irradiation dose, thereby improving fabrication tolerance and device uniformity.

The $\IcRn$ product also exhibited an exponential decay with increasing irradiation dose as shown in Fig.~\ref{fig:dose_dependence}(c).
The characteristic dose $\lambdazeroIcRn$ associated with the $\IcRn$ product showed a tendency to increase at lower temperatures (see Fig.~\ref{fig:dose_dependence}(d)).
Since the high-frequency performance of a Josephson junction is determined primarily by the characteristic voltage $V_{\mathrm{c}}=\IcRn$, operating at lower temperatures is advantageous from the viewpoint of reproducibly fabricating junctions with uniform properties.
The value obtained in the present study at \qty{3.6}{\kelvin}, namely \qty{90}{\ions/\nm}, is again in reasonable agreement with these earlier results.~\cite{Muller2019}
The normal-state resistance $\Rn$ increased exponentially with irradiation dose (see Fig.~\ref{fig:dose_dependence}(e)).
This behavior is also consistent with previous observations reported in the literature~\cite{Muller2019}.

\section{Discussion}
\subsection{Analysis based on SNS Josephson junction models}
Since the temperature dependence of $\Ic$ and $\IcRn$ reflects the characteristic energy scales governing Josephson transport, such as the superconducting gap, the proximity-induced minigap, and the Thouless energy, it serves as a sensitive probe of the transport mechanism.~\cite{Golubov2004}
In the following, we first analyze the behavior in the vicinity of $\Tc$ and then extend the discussion to the lower temperature regime based on theoretical models.

\begin{figure*}[t]
    \centering
    \includegraphics[width=0.32\textwidth]{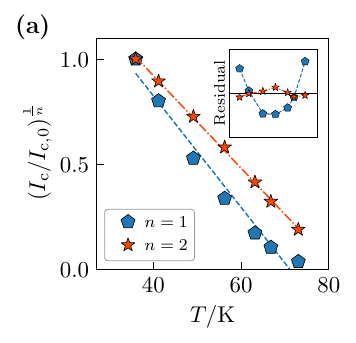}
    \includegraphics[width=0.32\textwidth]{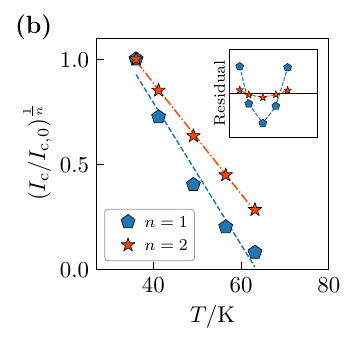}
    \includegraphics[width=0.32\textwidth]{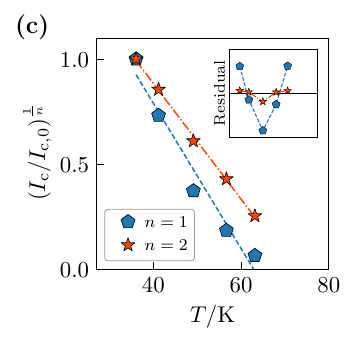}
    \caption{
            Plot of $(\Ic(T)/I_{\mathrm{c,0}})^{1/n}$ ($n=1,\,2$) for \qty{268}{\ions/\nm} (a), \qty{382}{\ions/\nm} (b), and \qty{421}{\ions/\nm} (c). $I_{\mathrm{c},0}$ is the critical current at the reference temperature of \qty{36}{\kelvin}. Broken and dash-dotted lines are fitting for $n=1$ and $n=2$, respectively. Inset: Fitting residuals.
            }
    \label{fig:degennes}
\end{figure*}
Figures~\ref{fig:degennes}(a)-\ref{fig:degennes}(c) show plots of $\Ic^{n}$ near $\Tc$ for $n=1$ and $n=2$.
According to the theoretical analysis by de Gennes~\cite{Gennes1964}, the critical current is expected to vary as $\Ic\propto(\Tc-T)^{n}$, with $n=1$ for SIS junctions and $n=2$ for SNS junctions containing a diffusive metallic interlayer in the dirty limit.
As is evident from those plots and fitting residuals (inset), the experimental data for $n=2$ is better fitted with a linear function.
This result suggests that the fabricated weak links exhibit behavior characteristic of dirty-limit SNS junctions rather than that of SIS junctions.
In particular, it suggests that the superconducting pair potential in the YBCO electrodes is reduced by the proximity effect near the weak-link region.
Similar behavior near $\Tc$, consistent with dirty-limit SNS-type junctions, has also been reported previously for YBCO weak links fabricated by helium-ion irradiation~\cite{Chen2022,Ogawa2024}.

Previous studies have shown that helium-ion irradiation of YBCO introduces disorder into the CuO chains, leading not only to a reduction in hole concentration but also to significant pair-breaking effects arising from enhanced electron scattering~\cite{Arias2003,Bobowski2010}.
By analogy, it is reasonable to assume that electrons experience strong scattering within the weak-link regions formed by focused helium-ion irradiation.
Motivated by these considerations, we model the superconductivity-suppressed weak-link region as a diffusive metal and analyze the low-temperature transport properties within the framework of dirty-limit SNS junction theory.
Furthermore, we examine whether the physical parameters extracted from this analysis are consistent with the expected properties of helium-ion-irradiated YBCO and assess the applicability of the dirty-limit SNS description to the present weak-link devices.

\begin{figure}[t]
    \centering
    \includegraphics[width=.95\columnwidth]{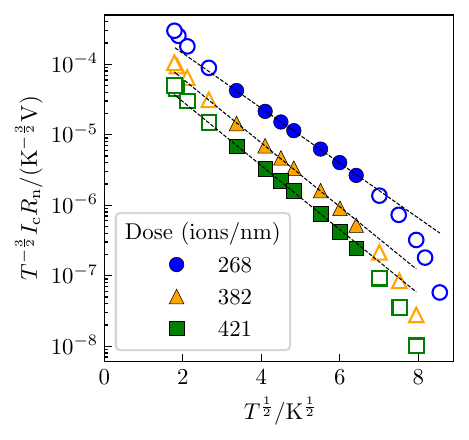}
    \caption{
                Curve fitting of the measured $\IcRn$ values using a dirty-limit long weak-junction model.
                Circles, triangles, and squares correspond to ion doses of \qtylist{268;382;421}{\ions/\nm}, respectively.
                Filled symbols indicate data points included in the fitting, while open symbols were excluded.
                The black dotted lines represent the model described by Eq.~\ref{eqn:fitting_model} in the main text.
            }
    \label{fig:thouless_fit}
\end{figure}
In \textit{d}-wave superconductors, the anisotropy of the superconducting pair potential is known to have a significant impact on junction properties through its influence on interfacial Andreev reflection.
Owing to the interplay and coexistence of the proximity effect and Andreev processes, Josephson junctions based on anisotropic superconductors exhibit complex transport behavior.~\cite{Tanaka1997}
A rigorous description of the transport phenomena would, in principle, require consideration of anisotropic material parameters, the angular dependence of the superconducting pair potential, scattering induced by interface barriers, and various forms of structural and electronic inhomogeneity~\cite{Yokoyama2007}.
However, it is difficult to uniquely determine all of these parameters from the transport data obtained in the present study.
Furthermore, since our primary objective is to provide a phenomenological description of the Josephson transport process in YBCO weak links, we intentionally employ simplified models that capture the essential physics while minimizing the number of adjustable parameters.
Accordingly, in the following analysis, we employ the theory of SNS junctions based on conventional \textit{s}-wave superconductors as an effective model to extract the dominant mechanisms governing the transport properties of the weak links.
Similar attempts of applying SNS-junction models to YBCO Josephson junctions has also been explored in previous studies.~\cite{Kabasawa1991,Sanchez-Manzano2022,Kirzhner2014aa}.

In the following, we follow the SNS-junction model presented in Ref.~\cite{Belzig1999}, which is based on the Usadel theory.~\cite{Usadel1970}
When the temperature satisfies the conditions $kT\gg{}\hbar{}D/L^{2}$ and $kT\ll{}\Delta$, the critical current of an SNS junction can be approximated by
\begin{align}
\Ic\simeq\frac{64\pi}{3+2\sqrt{2}}\frac{kT}{e\Rn}\frac{L}{\xi_{\mathrm{N}}}\exp\left(-\frac{L}{\xi_{\mathrm{N}}}\right),
\label{eqn:approx_model}
\end{align}
where $\xi_{\mathrm{N}}=\sqrt{{\hbar{}D}/{2\pi{}kT}}$ is the coherence length in the N region.
Here, $D$ denotes the electronic diffusion coefficient and $L$ is the junction length.
The quantity $\hbar{}D/L^{2}$ represents the characteristic energy scale associated with diffusive transport and is known as the Thouless energy $\ET$.
Using $\ET$, the ratio $L/\xi_{\mathrm{N}}$ can be rewritten as $L/\xi_{\mathrm{N}}=\sqrt{2\pi{}kT/\ET}$.
To fit our experimental data, we employed the following expression derived from Eq.~(\ref{eqn:approx_model}):
\begin{align}
\IcRn=aT^{\frac{3}{2}} \exp\left({-\sqrt{\frac{2\pi{}kT}{\ET}}}\right),
\label{eqn:fitting_model}
\end{align}
where $a$ and $\ET$ are the fitting parameters.
The coefficient $a$ is a phenomenological parameter introduced to account for factors that suppress the $\IcRn$ product.

In the present analysis, data points below \qty{10}{\kelvin} were excluded from the fitting procedure because in this temperature range, the system to fall outside the regime in which Eq.(\ref{eqn:fitting_model}) is valid.
In addition, data above \qty{45}{\kelvin} were also omitted because the superconducting energy gap in YBCO is not considered developed in this temperature range~\cite{Cybart2015}.

Figure~\ref{fig:thouless_fit} shows the results of the fitting analysis.
It can be seen that the transport properties of all junctions are well described by the SNS-junction model.
From these fits, the Thouless energy $\ET$ was extracted as \qtylist{0.67;0.50;0.50}{\meV} for the devices irradiated with doses of \qtylist{268;382;421}{\ions/\nm}, respectively (Fig.~\ref{fig:ET}(a)).
The corresponding temperatures $\ET$ are at most approximately \qty{7.8}{\kelvin} when expressed in temperature units.
Strictly speaking, the experimental temperature range does not well satisfy the symtotic condition $kT\gg{}\ET$.
For example, at $T=\qty{10}{\kelvin}$, we have $kT/\ET\simeq 1.28$. The contribution of the second Matsubara frequency is then estimated to be approximately \qty{22}{\percent} of the lowest-frequency contribution.
Nevertheless, we use the lowest-frequency expression Eq.(\ref{eqn:approx_model}) as a analytical leading-order description of the temperature dependence.
Since the present analysis concerns only the relative temperature dependence rather than the absolute magnitude, uncertainties that enter predominantly through the overall prefactor are reasonably neglected.

In contrast to the exponential dose dependences observed for $\Ic$, the $\IcRn$ product, and $\Rn$, the variation of $\ET$ is relatively modest.
For SNS junctions with highly transparent interfaces, $\ET$ is known to be the sole energy scale governing the $\IcRn$ product.~\cite{Golubov2004,Dubos2001,Belzig1999}
Consequently, in the low-temperature limit, the dimensionless ratio $e\IcRn/\ET$ approaches a universal value of approximately $10.82$.
As shown in Fig.~\ref{fig:ET}(b), however, the values of $e\IcRn/\ET$ obtained for the weak links investigated in this work are $2.7$, $1.2$, and $0.57$ for the \qtylist{268;382;421}{\ions/\nm}, respectively.
These values are approximately one order of magnitude smaller than those expected for highly transparent SNS junctions.
Furthermore, the ratio decreases systematically with increasing irradiation dose.
This result indicates that the pronounced suppression of $\IcRn$ cannot be explained solely by the reduction of the Thouless energy.
Additional mechanisms, such as a decrease in interface transparency, enhanced pair-breaking, or a reduction in the effective number of conducting channels participating in Josephson transport, are likely to contribute.
The possible origins of this behavior are discussed in the following section.

It is worth noting that the effective diffusion coefficient $D$ within the weak-link region can be estimated from the extracted effective Thouless energy using the microscopic relation $\ET=\hbar{}D/L^{2}$.
Taking the effective junction length of the \qty{268}{\ions/\nm} device to be $L=\qty{15}{\nm}$, as estimated from the measured value of $\Rn$, an effective diffusion coefficient is $D=\qty{2.3}{\cm\squared/\s}$.
This estimation is approximately one order of magnitude smaller than the value of $D=\qty{20}{\cm\squared/\s}$ reported in a previous study on YBa${}_{2}$Cu${}_{3}$O${}_{6.5}$.~\cite{Gedik2003}
Such a reduction is reasonable considering that helium-ion irradiation is expected to introduce substantial scattering centers into the weak-link region.
Using this value of $D$, the coherence length in the normal region $\xi_{\mathrm{N}}$ is estimated to be approximately \qty{5.3}{\nm} at \qty{10}{\kelvin}, which is one-third as long as the estimated $L$.

\subsection{Phenomenological description of the YBCO weak links}
\begin{figure*}[t]
    \centering
    \includegraphics[width=.32\textwidth]{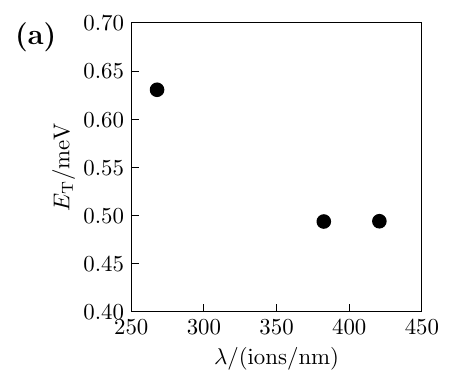}
    \includegraphics[width=.32\textwidth]{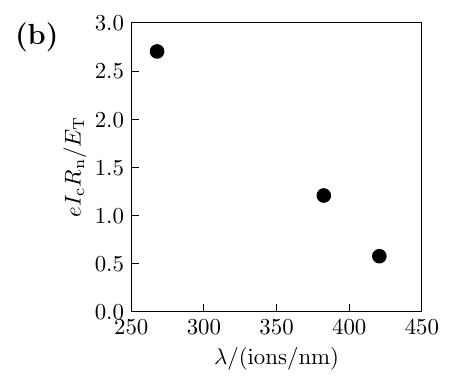}
    \includegraphics[width=.32\textwidth]{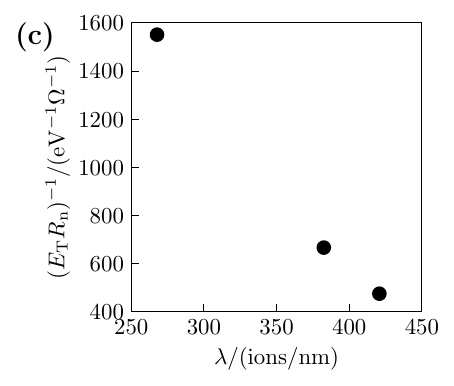}
    \caption{
            Dose ($\lambda$) dependence of effective Thouless energy $\ET$ (a), $\IcRn/\ET$ (b), and $1/(\ET\Rn)$ (c).
            }
    \label{fig:ET}
\end{figure*}
In the following, we examine the dose dependences of the Thouless energy $\ET$, the $\IcRn$ product, and the normal-state resistance $\Rn$, and consider the physical picture that emerges from these observations.
As noted previously, the ratio $e\IcRn/\ET$ exhibits a clear dependence on irradiation dose in YBCO weak links (see Fig.~\ref{fig:ET}(b)).
This behavior suggests that the $\IcRn$ product is not determined solely by $\ET$.
Within the framework of SNS-junction theory, this can be interpreted as a consequence of the suppression of $\IcRn$ by the boundary conditions at the SN interfaces, with the suppression becoming increasingly pronounced as the irradiation dose increases.

One of the most common mechanisms leading to a reduction of $\IcRn$ in SNS junctions is a decrease in interface transparency.~\cite{Hammer2007}
Defect layers or insulating regions formed by oxidation at the interface between the superconducting electrodes and the normal interlayer give rise to an interfacial barrier potential $Z$, reducing the electronic transmission probability across the interface and thereby strongly suppressing $\IcRn$~\cite{Ting2016,Bubis2017}.
In addition, the effective interface transparency may also be reduced by a mismatch in Fermi velocity between the superconducting (S) and normal (N) regions~\cite{Tuuli2012,Daghero2011}.

Furthermore, the anisotropic nature of the \textit{d}-wave superconducting order parameter may further suppress the pair amplitude induced in the N region through the proximity effect, leading to an additional reduction of $\IcRn$.~\cite{Sigrist1992,Yokoyama2007}
In YBCO weak links formed by helium-ion irradiation, the SN interfaces are unlikely to be perfectly specular; instead, the irradiation-induced disorder is expected to create local interface structures with random orientations relative to the YBCO crystal lattice.

As a consequence, a finite effective misorientation is expected to exist between the junction interface and the crystallographic axes of YBCO, which can further suppress the $\IcRn$ product.~\cite{Golubov2004}
Another possible origin of the suppression of the $\IcRn$ product is a reduction in the effective density of states participating in Andreev reflection at the SN interfaces.
If helium-ion irradiation reduces the electronic density of states in YBCO, the effective number of transport channels available for Andreev processes may likewise be reduced, leading to a suppression of the Josephson coupling strength.
As discussed in the following paragraph, a reduction in the density of states also provides a plausible explanation for the observed increase in the normal-state resistance $\Rn$.
Therefore, changes in the density of states induced by helium-ion irradiation may offer a unified interpretation of both the suppression of $\IcRn$ and the enhancement of $\Rn$.

We next consider the relationship between the obtained effective Thouless energy $\ET$ and the normal-state resistance $\Rn$.
Using the Einstein relation for electrical conductivity,
$\sigma=2N(0)De^{2}$ the normal-state resistance can be written as
\begin{align}
\Rn=\frac{L}{\sigma{}A}=\frac{1}{2AN(0)e^{2}}\frac{L}{D},
\end{align}
where $A$ is the cross-sectional area of the junction and $N(0)$ is the effective local density of states participating in transport.
As we saw in Fig.~\ref{fig:temp_dependence}(e), $\Rn$ increases exponentially with increasing irradiation dose, whereas $\ET=\hbar{}D/L^{2}$ exhibits only a weak decreasing trend (see Fig.~\ref{fig:ET}(a)).
Those observed behaviors suggest that the increase in $\Rn$ cannot be primarily attributed to changes in either the diffusion coefficient $D$ or the effective junction length $L$.
Instead, it can be naturally explained by a reduction in the density of states $N(0)$ contributing to electronic transport.
Indeed, the quantity $(\ET\Rn)^{-1}=2(e^{2}/\hbar)N(0)AL$ shows a monotonic decline with increasing helium ion irradiation dose. (see Fig.~\ref{fig:ET}(c))
Since the junction cross-sectional area $A$ is independent of irradiation dose, and the effective junction length $L$ is expected to increase rather than decrease with increasing dose, this trend strongly suggests a reduction in the effective density of states $N(0)$ within the irradiated weak-link region.

Although the microscopic mechanism responsible for such a reduction in the local effective density of states in YBCO weak links cannot be identified conclusively at present, the observed behavior is consistent with the suppression of the density of states associated with the pseudogap phase universaly observed in underdoped cuprate high-temperature superconductors, including YBCO~\cite{Ding1996aa,Hashimoto2014aa,Fradkin2015,Yokoyama2016}.
These observations suggest that the introduction of disorder on the nanoscale may induce a pseudogap-like reduction of the density of states.

Finally, we discuss the implications of the present work.
In this study, the transport properties of YBCO weak links were described phenomenologically in terms of relationships among physically interpretable parameters.
This approach provides a framework for Josephson-junction engineering based on underlying physical mechanisms rather than empirical optimization alone.
Moreover, the diffusive description characterized by an effective diffusion coefficient $D$ may offer design principles for controlling helium-ion irradiation patterns from the perspective of carrier diffusion.
In particular, nanoscale modulation of the irradiation dose may provide a route toward further improvements in device performance and uniformity.

The present study also provides a general framework for utilizing Josephson transport as a probe of nanoscale electronic states.
In general, for SNS-type Josephson weak links, it is difficult to directly interpret $\mathrm{d}I/\mathrm{d}V$ as the density of states, unlike in conventional tunneling spectroscopy.
The quantity $(\ET\Rn)^{-1}$ employed in the present study can serve as a useful metric for evaluating the effective number of low-energy states contributing to transport.
This concept is not limited to YBCO, but may also be extended to conductive weak links with locally modulated electronic states, including topological insulators/superconductors, semiconductor-superconductor hybrids, magnetic metals, strongly correlated metals, and phase-change materials.

\section{Conclusion}
We have investigated the transport properties of helium-ion-microscopy-fabricated \YBCO weak links with irradiation doses of \qtylist{268;382;421}{\ions/\nm} and analyzed the results within the framework of dirty-limit SNS Josephson junction theory.
The temperature dependences of both the critical current $\Ic$ and the characteristic voltage $\IcRn$ were consistently described by a diffusive SNS model over a wide temperature range, suggesting that Josephson transport in the irradiated weak-link region is governed predominantly by diffusive proximity coupling.

Analysis of the experimental data yielded an effective Thouless energy $\ET$ of approximately \qtyrange[range-phrase={--}]{0.5}{0.67}{\meV} with only a weak dependence on helium-ion dose.
In contrast, both $\Ic$ and $\IcRn$ decreased strongly with increasing dose, while $\Rn$ increased nearly exponentially.
These trends cannot be explained solely by variations in $\ET$, the diffusion coefficient, or the effective junction length.

The observed behavior is instead consistent with a decrease in interface transparency and the effective number of conducting channels participating in Andreev refrection at the interface arising from a decrease of the local density of states within the irradiated region.
The decrease of the local density of states naturally accounts for the increase of $\Rn$.

The present work provides a consistent phenomenological framework for understanding transport in helium-ion-irradiated YBCO weak links.
Beyond its relevance to Josephson junction engineering, the approach demonstrates how transport measurements can be used to extract information on nanoscale electronic properties in locally modified cuprate superconductors.
The framework may also be applicable to a broader class of weak-link systems in which disorder or local electronic-state engineering governs superconducting transport.

\section*{ACKNOWLEDGMENTS}
The authors are grateful to Tomohiko Iijima for his assistance with the helium ion microscopy (HIM) fabrication process. This work was supported by JSPS KAKENHI Grant Numbers JP20H02631, JP23H01458, 23K26152, and JP26K17438. A part of this work was supported by "Advanced Research Infrastructure for Materials and Nanotechnology in Japan (ARIM)" of the Ministry of Education, Culture, Sports, Science and Technology (MEXT). Proposal Number JPMXP1225AT0161.

\section*{data availability}
The data that support the findings of this study are available from the corresponding author upon reasonable request, subject to institutional policies and restrictions.

\bibliography{refs} 
\end{document}